\documentclass[twocolumn,10pt]{IEEEtran}

\usepackage[ruled,vlined]{algorithm2e}
\usepackage{pgfplots}
\usetikzlibrary{backgrounds,automata}
\usetikzlibrary{plotmarks}
\usetikzlibrary{patterns}
\usetikzlibrary{calc,positioning,fit,backgrounds}
\usetikzlibrary{shapes,snakes}
\usetikzlibrary{intersections,positioning}
\usepackage{caption}
\usepackage{cite}
\usepackage{amsmath,amssymb,amsfonts,color,soul,amsthm}
\usepackage{algorithmic}
\usepackage{graphicx}
\usepackage{textcomp}
\usepackage{xcolor}
\include{notation}
\usepackage[caption=false,font=footnotesize]{subfig}
\def\BibTeX{{\rm B\kern-.05em{\sc i\kern-.025em b}\kern-.08em
    T\kern-.1667em\lower.7ex\hbox{E}\kern-.125emX}}

\begin{document}

\title{SDN assisted UAV communication systems : Efficient Deployment Strategies

}
\author{
\IEEEauthorblockN{Sai Teja Suggala, Siddhartha Pothukuchi, and Naimat Ali Khan}\\
\IEEEauthorblockA{\textit{Dept. of Electrical Engineering and Computer Science \\
Indian Institute of Technology Bhilai, India }\\
Email: \{suggalat, pothukuchis, naimatk\}@iitbhilai.ac.in}
}

\maketitle

\begin{abstract}
Recently, Unmanned Aerial Vehicle (UAV) based communications systems have attracted increasing research and commercial interest due to their cost effective deployment and ease of mobility.During natural disasters and emergencies, such networks are extremely useful to provide communication service. In such scenarios, UAVs position and trajectory must be optimal to maintain Quality of Service at the user end. This paper focuses on the deployment of an SDN-based UAV network providing communication service to the users. We consider the deployment of the system in stadiums and events. In this paper, we propose a scheme to allocate UAVs to the users and a traffic congestion algorithm to reduce the number of packets dropped to avoid re-transmissions from the user end. We also propose an energy efficient multi hop routing mechanism to avoid the high power requirement to transmit longer distances. We assume that all the back-haul links have sufficient capacities to carry all the traffic from the front-haul links and the design of UAVs must consider their power requirements for both flight and transmission.

\end{abstract}


\section{Introduction}

The static nature of any terrestrial base stations limits their ability to handle traffic surges in stadiums, events, emergency situations where infrastructure is compromised. Recently, Unmanned Aerial Vehicle (UAV) based communications systems have attracted increasing research and commercial interest. In such networks, UAVs carry the equipment, such as a base station to provide communication service. UAV-based communication networks are extremely useful to provide communication service in any compromised network scenario. The important aspects of UAV based communications that attract huge attention are high-quality air to ground links (LoS links), cost-effectivity, fewer penetration losses at back-haul and better QoS of 5G RANS.

UAV Communication system consists of two types of links, which are front-haul and back-haul for UAV-User and UAV-SDN communication respectively. Usually, the backhaul links are dominant due to the high altitude deployment of UAVs that can avoid high penetration losses on mmWave bands except for the multipath and rain attenuation of 10-30 dB.

On the network infrastructure side, Software Defined Networking (SDN) has emerged as a promising alternative to the traditional IP networks. SDN consists of a central controller and forwarding elements (switches and routers) and it separates the network logic (control plane) from the data plane. This centralized control can make SDN ideal in a UAV network due to the size of UAVs and the large area it is deployed in. All the nodes i.e., UAVs transmit their placement, network load and other statistics periodically to the controller. After gathering the network statistics from the nodes, SDN installs flows on the nodes. Thus, the involvement of SDN can make the UAV communication more efficient than any deployment setup. 

\begin{figure}[htbp]
\centerline{\includegraphics[width=8cm]{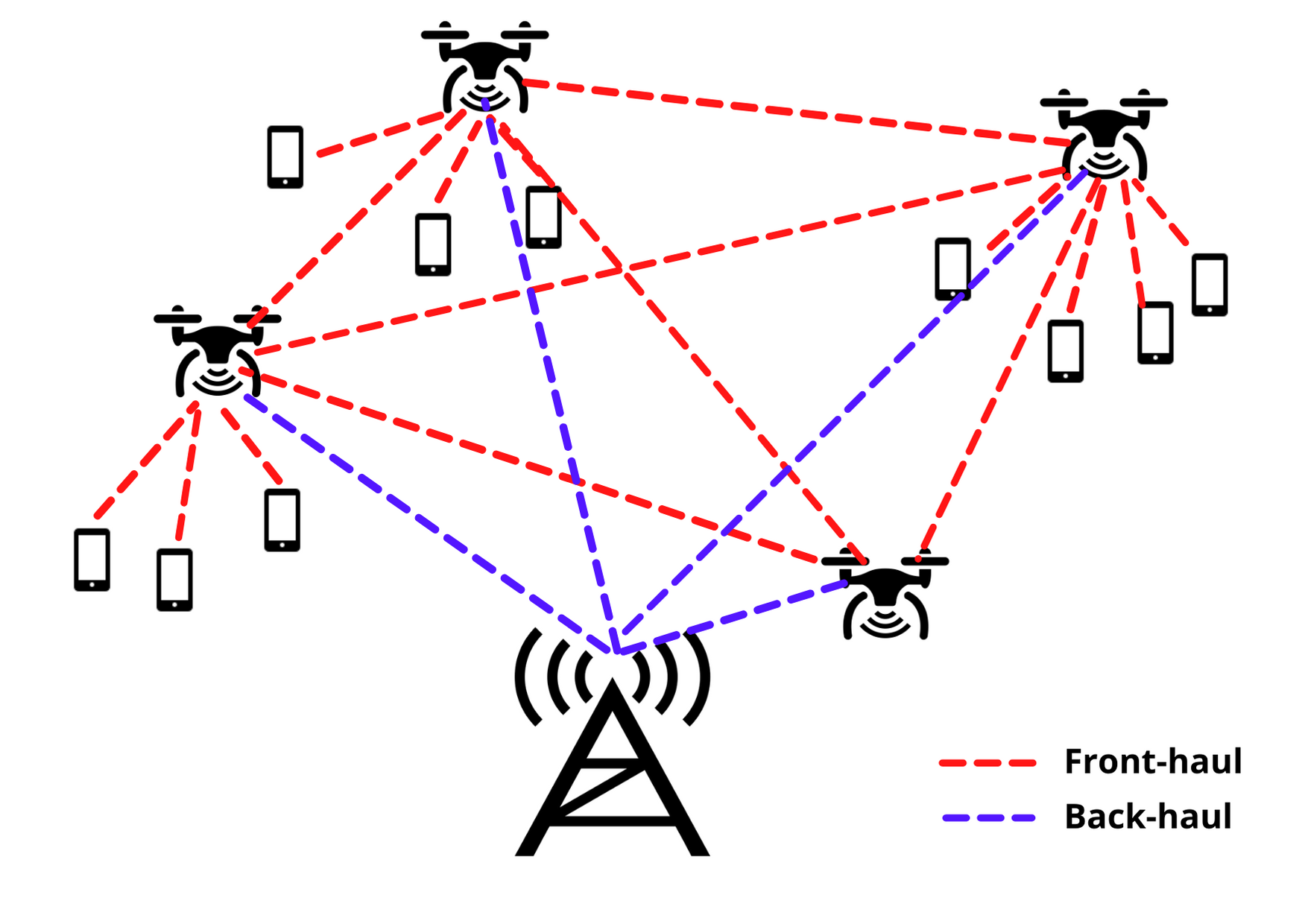}}
\caption{Network Deployment with UAVs, user devices and SDN controller.}
\label{fig_setup1}
\end{figure}

The main challenges of such a system are resource allocation to the users, UAV trajectory, placement optimization, user association, limited buffer size and limited power at UAVs to transmit and flight. The main focus will be on the placement of such nodes to serve traffic surges and to implement an algorithm to tackle traffic congestion. This algorithm can reduce the number of packets dropped at the UAVs. We also consider multihop communication from UAVs to the controller to reduce the power required to transmit for long distances. As in the figure \ref{fig_setup1}, the closest UAV to the controller is considered as the root node that only forwards the UAV traffic to the controller.

\section{Related Work}

Placement and Trajectory Optimization has grabbed a lot of attention in the research community, with a major problem of non-uniform traffic congestion. Optimization of the locations of UAVs and channel assignment are jointly investigated in \cite{ZOU2020} by exploiting partially overlapped channels (POCs). In \cite{8038869}, a novel framework has been proposed for efficient deployment and UAV mobility to collect data in the uplink from ground Internet of Things (IoT) devices. The results have shown that by intelligently moving and deploying the UAVs, the total transmit power of the devices significantly decreases compared to the case with pre-deployed stationary aerial base stations. Power consumption of UAVs for their flight and transmission is extremely important. Maximizing the battery life by optimizing the placement of UAV has been studied in \cite{8450437}. In crowded areas, bandwidth allocation is very crucial in order to maintain Quality of Service considering all the 3 types of services i.e., control, real-time and non-real-time communications. \cite{9120678} shows that the performance of the system can be enhanced by clustering the user groups and allocating each cluster a unique RF channel with different bandwidth. An alternative optimization approach of UAV locations based on their LoS/NLoS probabilities and achievable rates to optimize the bandwidth allocation and UAV placement until convergence is studied in \cite{9014076}.
The exchange of control packets is crucial for the operation of the SDN-based UAV network. \cite{8190924} focuses on controller placement and the trade-off between control overhead and delay. Merging of control packets to reduce overhead is proposed in multi-hop communication to reduce the power required to transmit to the faraway controller. 

In \cite{7569080}, a novel strategy was proposed for the allocation of UAVs to a demand area using entropy nets where the problem was formulated as a minimax facility problem and optimizing placement came out to be the viable solution. Simulation results proved the approach was capable of optimizing the allocation delays without affecting the network's capacity and coverage. In \cite{7486987}, the downlink coverage probability for UAVs as a function of the altitude and the antenna gain is derived then using circle packing theory, optimal 3-D locations are determined considering maximum coverage area. 

When there is a constraint on the number of UAVs to be used, the number of packets will increase extremely resulting in more re-transmissions from the User end. This aspect is solved in \cite{7486987}, where the user gets feedback from the base station to the user via the RLC layer to reduce the traffic generation rate at the users. This avoids queue overflow at the UAV, but this concept must be used only when there is a constraint on the number of UAVs to be deployed since reducing the traffic generation rate affects the quality of service.

\section{System Model}

We consider a specific network area such as stadiums where the user devices exist between the stadium and the ground. The UAV based wireless network is integrated to provide coverage for all the user devices. We define $M$ as the set of UAVs in the network, $M$ = {1, ..., m} and $N$ as the set of all UEs, i.e., $N$ = {$1,2,3,.., n $}, and let $N_{j}$ denote the set of UEs associated with UAV $j$. We consider the altitude of flight is constant for all the UAVs, so coordinates of the UAV are $R_{j}^{d}$ = ($x_{j}^{d} , y^{d}_{j} , H$ ), $\forall$j $\in$ M. Moreover, UEs are assumed to have coordinate $R_{i}^{d}$ = ($x_{i}^{d} , y^{d}_{i}$) , $\forall$i $\in$ $N$. We define the coordinates of the cellular Base station as  ($x^{b}, y^{b}$) which is a gateway to the EPC Core and also acts as SDN controller. Backhaul Link uses mmWave( Many studies have shown that with frequencies between $50$GHz- $100$GHz the LoS transmission can go up to $6-8$km. Access link also uses mmWave-(Beam Steering Technology that allows the massive MIMO base station antennas to direct the radio signal to the users rather than in all directions, providing less latency and higher throughput.)

\begin{figure}[htbp]
\centerline{\includegraphics[width=8cm]{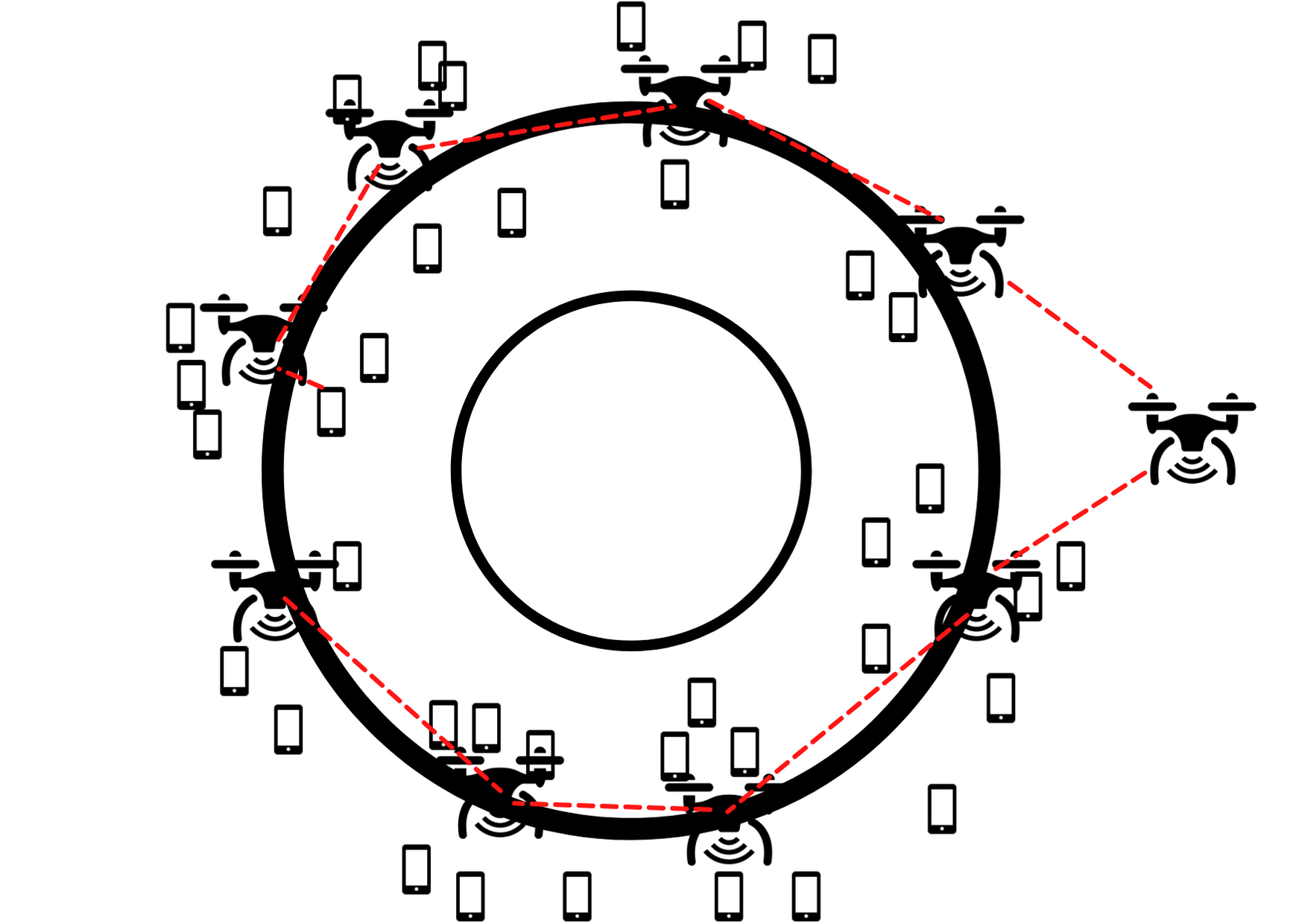}}
\caption{Depicts system topology in a stadium scenario and the best path for multi hop routing between the root node and each UAV. }
\label{systemtopology}
\end{figure}

The root node in the Fig. \ref{systemtopology}, is meant to collect traffic from UAVs and forward to the controller. The root node does not serve any user devices and is of larger size since it receives higher amount of traffic from all the other UAVs and the backhaul (UAV - Controller) of the root node is of high quality LoS link achieving higher channel throughput. To avoid Single Point of Failure (SPOF) at root node, the other UAVs must elect a root node using heart beat protocol. The heart beat is a periodic signal generated by each UAV to synchronize with other UAVs at a regular intervals. The protocol is generally used in UAV communication to negotiate and monitor the SPOF possibility, an election process with other UAVs on heart beat network to determine which UAV can act as the root node.

\section{Proposed Work}
Our main objective is to make the SDN assisted UAV communication systems more efficient. By optimizing placement locations of UAVs, traffic congestion can be reduced. By the implementation of multi-hop routing, the power required for transmission decreases by ten folds resulting in higher flight times i.e., energy efficient technique. The objectives of our proposed work are briefed as follows. 

\begin{itemize}
    \item Minimum Number of UAVs required to serve all the user devices.
    \item An efficient User-UAV allocation algorithm based on traffic requirements. 
    \item Placement Updation of UAVs to serve traffic surges and improve QoS.
    \item Implementation of traffic congestion algorithm proposed in \cite{7486987}.
    \item An efficient routing path between root node and UAVs for multi hop communication.
\end{itemize}

 \begin{algorithm}[]
\SetAlgoLined
 Segmentation()--$>$ Stadium divided into sectors based on range of UAV\;
 AllocUAV()--$>$ Allocate UAVs to respective sectors and decide number of UAVs required.\;
 UserToUav()--$>$ Allocate users to UAVs
 \While{ True}{
  UpdateTraffic()--$>$ Update traffic at each BS\;
  Placement()--$>$ Update UAVs placement based on user traffic\;
  TrafficCongestionControl()\;
  MultiHopRouting()\;
 }
 \caption{System Flow}
\end{algorithm}

\subsection{Minimum number of UAVs}

We divide the whole circular deployment into $S$ number of sectors using the maximum range of coverage ($R$) of each UAV. As in Fig. \ref{maxrange}, User located in the range $[ R'-R, R'+R ]$  can get served by a UAV, since the maximum coverage of the UAV is $R$. We can find the maximum number of sectors that can be obtained with a maximum range of R as follows. Coordinates of user devices are assigned in the specified range of $[ R'-R, R'+R ]$ randomly as in Fig. \ref{usercircle}. 

Based on the user's angle with the center of the circle, mapping of each user with sectors can be done to form user groups. Based on this grouping, we can obtain the number of UAVs required by finding the traffic requirements and the number of users in a certain segment.

Let $T(S_n)$ be the traffic generated in a particular interval in the $n$th sector and $B$ be the maximum buffer capacity of one UAV. So, the number of UAVs ($N(S_n)$) required in $S_n$ can be written as  
\begin{equation}
    N(S_n)=T(S_n)
\end{equation}

From this, we can obtain the initial number of UAVs required, but this may not be the minimum number of UAVs required. By checking traffic requirements periodically, UAVs can be added/removed from their assigned sector. This scenario is explained in the latter sections of this paper.

\begin{figure}[htbp]
\centerline{\includegraphics[width=8cm]{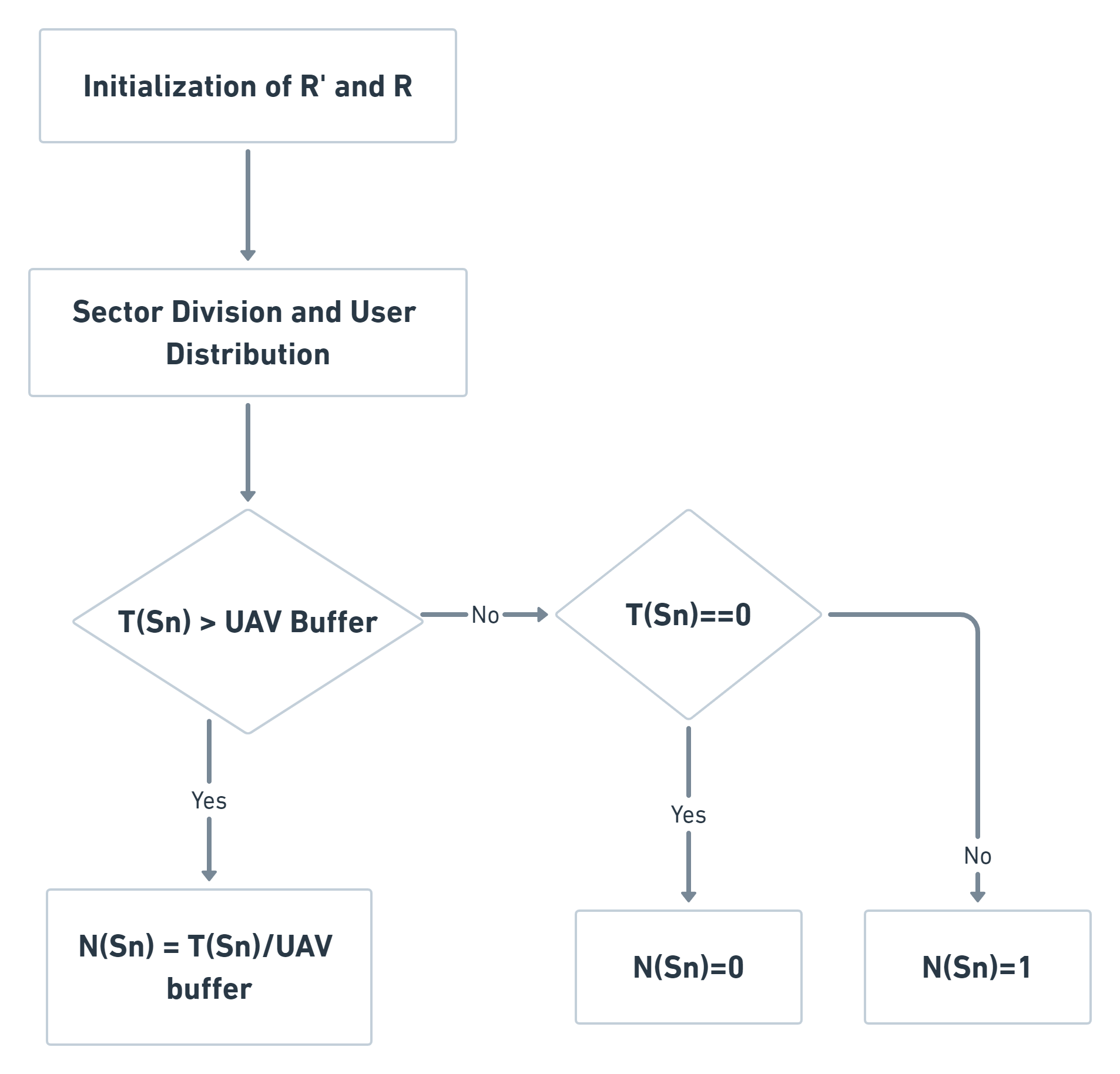}}
\caption{ Flow Chart to find the number of UAVs required in a sector. }
\label{maxrange}
\end{figure}

\begin{figure}[htbp]
\centerline{\includegraphics[width=8cm]{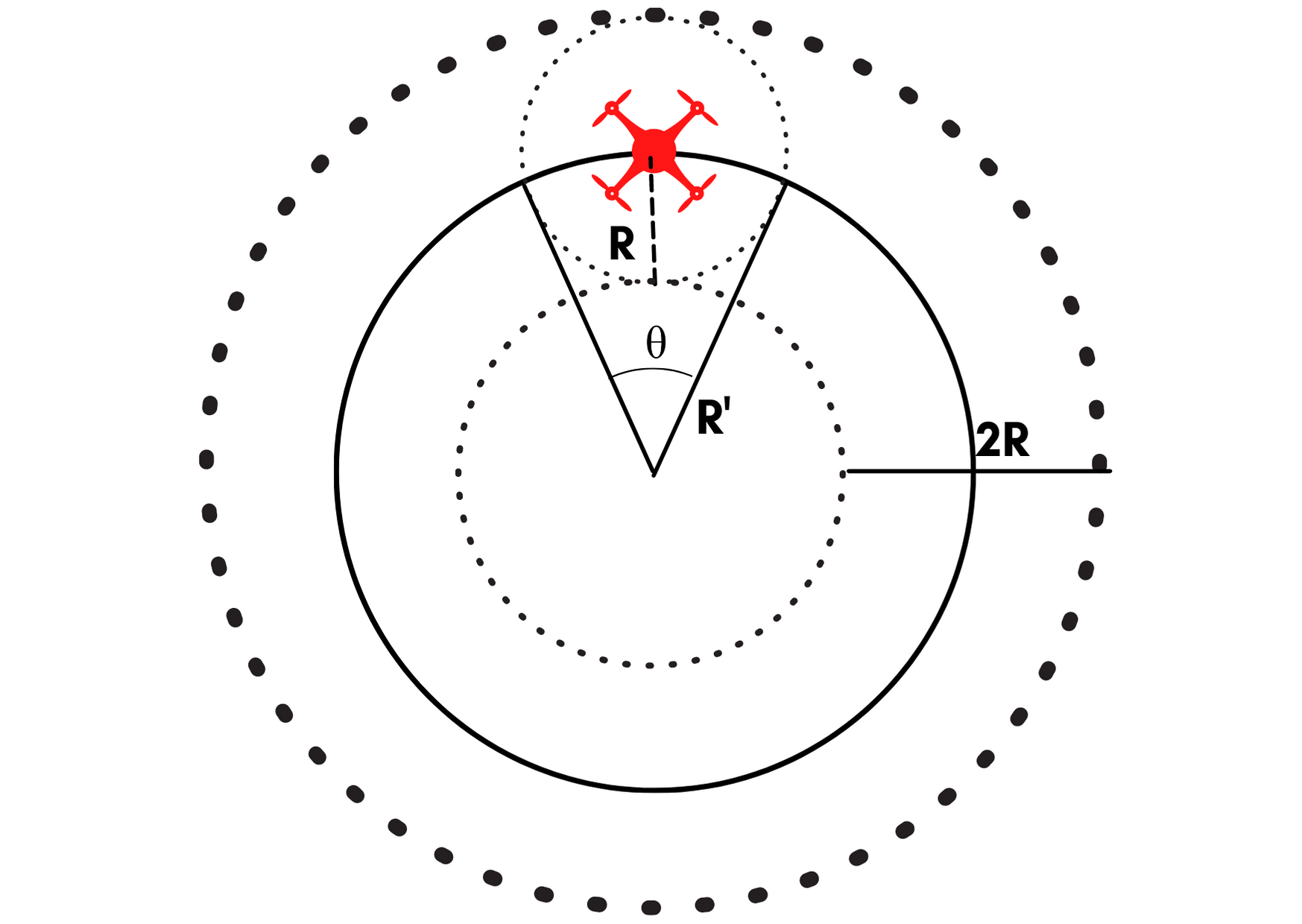}}
\caption{ Representation of sectors, maximum range of UAV  and possible user coverage. }
\label{maxrange}
\end{figure}

\begin{figure}[htbp]
\centerline{\includegraphics[width=8cm]{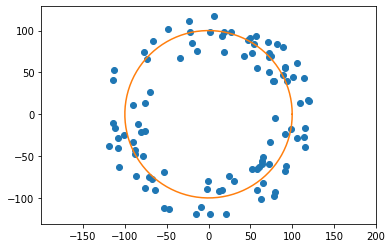}}
\caption{User distribution for radius $(R')$=$100$ units, $R$=$25$ units and No. of Users $(N)$=$100$}
\label{usercircle}
\end{figure}

\begin{figure}[htbp]
\centerline{\includegraphics[width=8cm]{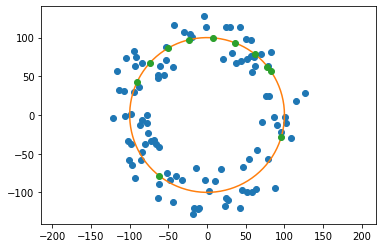}}
\caption{UAV allocation for radius $(R')$=$100$ units, $R$=$25$ units and No. of Users $(N)$=$100$}
\label{usercircle}
\end{figure}

\subsection{UAV Placement}
User-UAV traffic rate and their type i.e., real-time/non-real-time are captured frequently to evaluate the traffic requirements of each sector. A Scoring algorithm based on the type of traffic and traffic rate is proposed to rank the sectors. Based on this ranking, neighbouring UAVs that are free or partially free will join this sector or change their location nearer to the congested sector. This way, neighbouring UAVs can serve at least a few user devices of the congested sector to reduce the load in the congested sector.

\begin{figure}[htbp]
\centerline{\includegraphics[width=8cm]{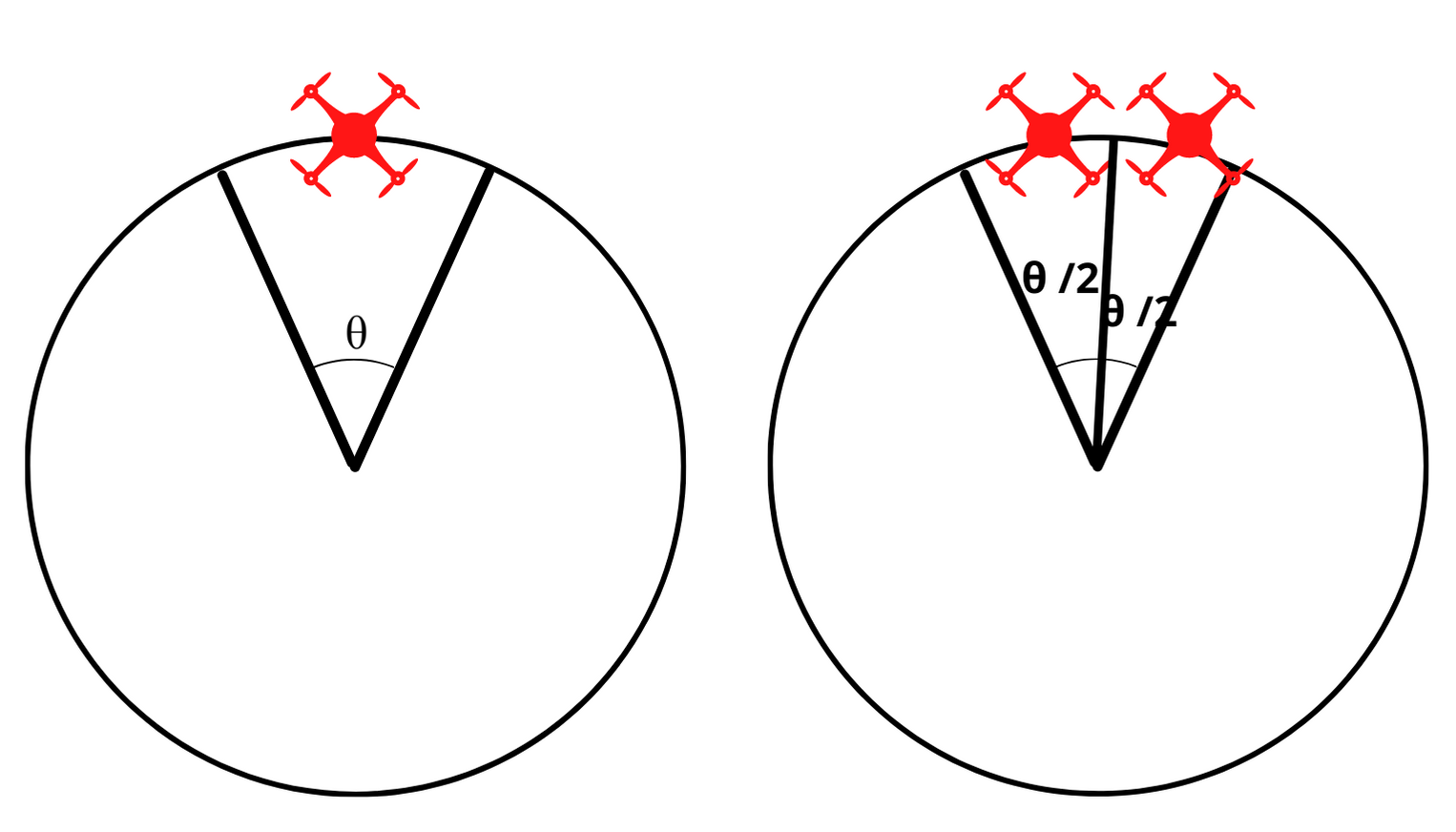}}
\caption{Distribution of UAVs in a sector}
\label{placement}
\end{figure}

As in fig.\ref{placement} We follow the symmetrical placement of UAVs in the sector to distribute users among the serving UAVs in a certain sector. Traffic surges can be tackled by implementing a moving average algorithm on the traffic received at UAVs. UAVs must send this information to the controller to predict the traffic surge and accordingly update the positions of the UAVs in each sector.

 \begin{algorithm}[]
\SetAlgoLined
 UAVlist is the list of all users a UAV serves\;
  \If{UAV==busy}{
   \eIf{neighbourUAV==busy}{
   Trafficcongestioncontrol() or sendnewUAV()\;}
   {
   moveneighbourUAV()\;}
   }
 \caption{UAV Placement}
\end{algorithm}

A new UAV must be deployed when every UAV is congested and the UAV placement is updated to allocate UAVs in the new topology. When every UAV is either free or partially occupied, a set of UAVs must be suspended until any traffic rise. Periodic transmission of network topology, link capacities and traffic patterns to the controller is crucial to maintain the system stable.

\subsection{Traffic Congestion Control}
Packet Overflow occurs whenever the buffer capacity has been utilized i.e., high traffic generation rate at the user. In this scenario, a new UAV can be deployed to distribute the traffic receiving at UAVs. When there is a constraint on the number of UAVs to be deployed, then it is very difficult to control the traffic congestion. To the best of our knowledge, no research work is done on this scenario. So, we consider two types of buffer capacities for UAV-UAV and User-UAV routing. We implement the traffic congestion technique proposed in \cite{7486987}, only when there is a constraint on the maximum number of UAVs.

\begin{figure}[htbp]
\centerline{\includegraphics[width=8cm]{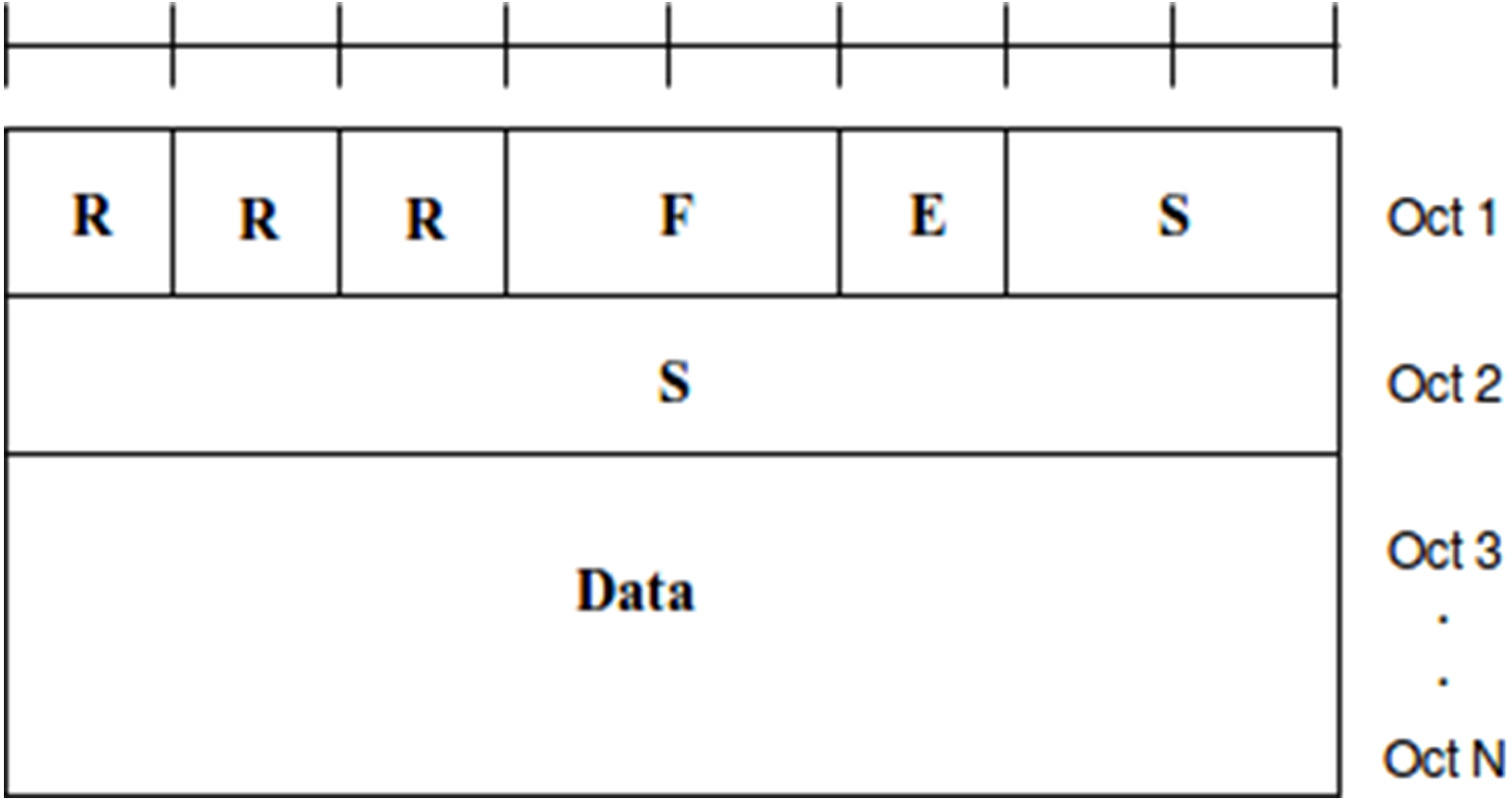}}
\caption{The First 3 bits (R s) of the RLC header.}
\label{rlc}
\end{figure}

As in Fig. \ref{rlc}, feedback from the base station to the user device can be done using the first 3 bits of the RLC header which are reserved and unused. Thus, the base station can send feedback to reduce the traffic generation rate by $k$ units where $T(N_i)$ is the current traffic generation rate of the user $i$:

\begin{equation}
    k=max(UAVbuffer)/T(N_i)
\end{equation}

 \begin{algorithm}[]
\SetAlgoLined
 UAVlist is the list of all users a UAV serves\;
 \For{ i in len(UAVlist)}{
  instructions\;
  \If{priority==Realtime}{
   usertraffic=usertraffic*0.9\;
   }

  \If{priority==nonRealtime}{
   usertraffic=usertraffic*0.8\;
 }
 \If{priority==Control}{
   usertraffic=usertraffic*1\;
 }
 }
 \caption{Traffic Congestion Control}
\end{algorithm}

\begin{figure}[htbp]
\centerline{\includegraphics[width=9cm]{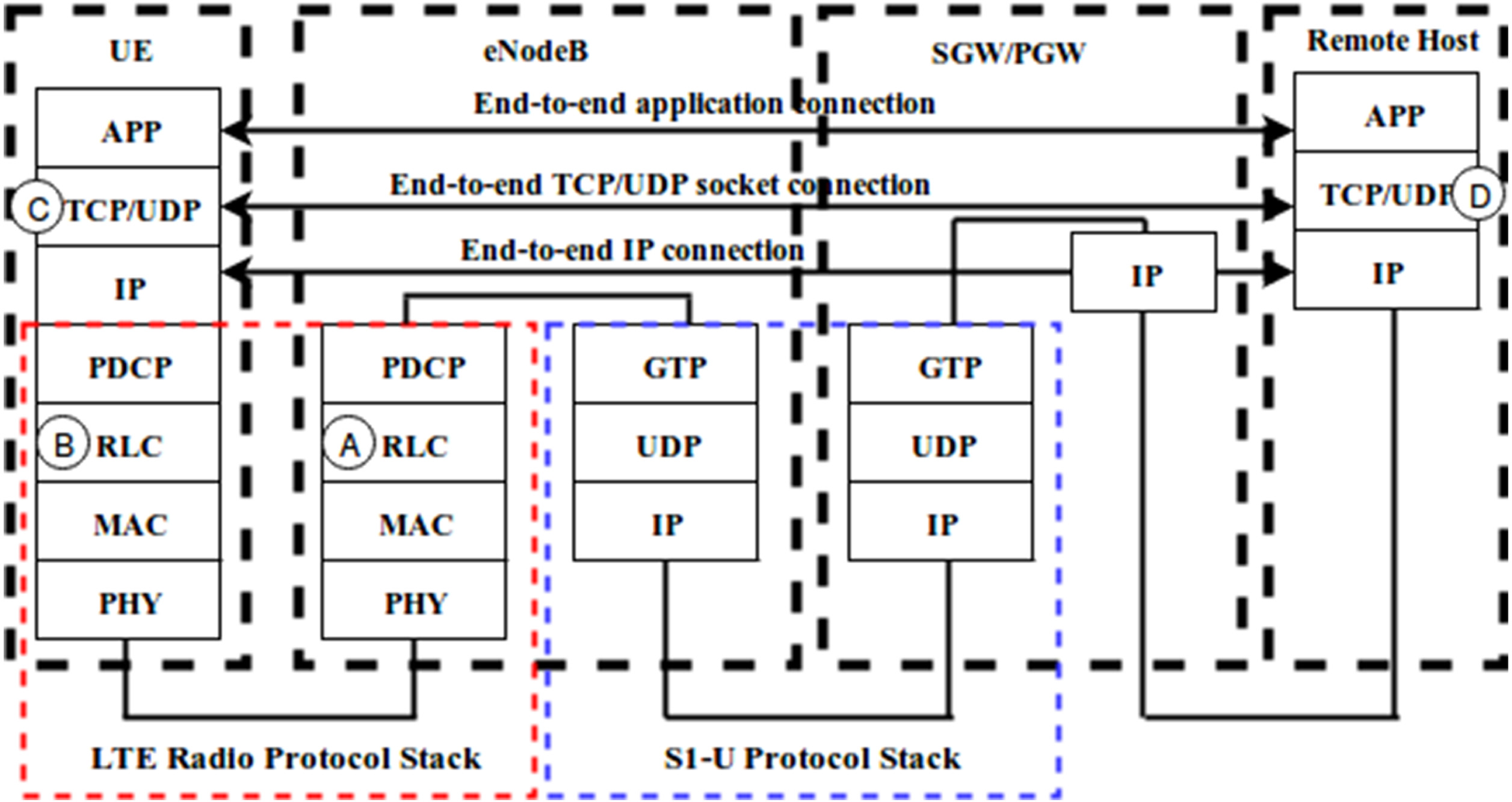}}
\caption{Flow of congestion information from the base station to the sender.}
\label{congestionflow}
\end{figure}

\subsection{Multi Hop Communication with Controller}
Multi-Hop Communication can make the system energy efficient by reducing the need to transmit long distances. In order to implement multi-hop communication, the sender must know the least number of hops required to send the data to the controller. So, we implement Dijkstra's algorithm to find the shortest path to send the packets to the controller, since Dijkstra's algorithm is weighted and guarantees the shortest path. We need to consider the network load on the receiver and the distance between the sender UAV and receiver UAV. So, we consider the linear combination of the network load and the distance between the sender UAV and receiver UAV as the weight of the edge in Dijkstra's algorithm. We obtain a minimum spanning tree, via which all the hops are to be routed to the controller.

\begin{figure}[htbp]
\centerline{\includegraphics[width=8cm]{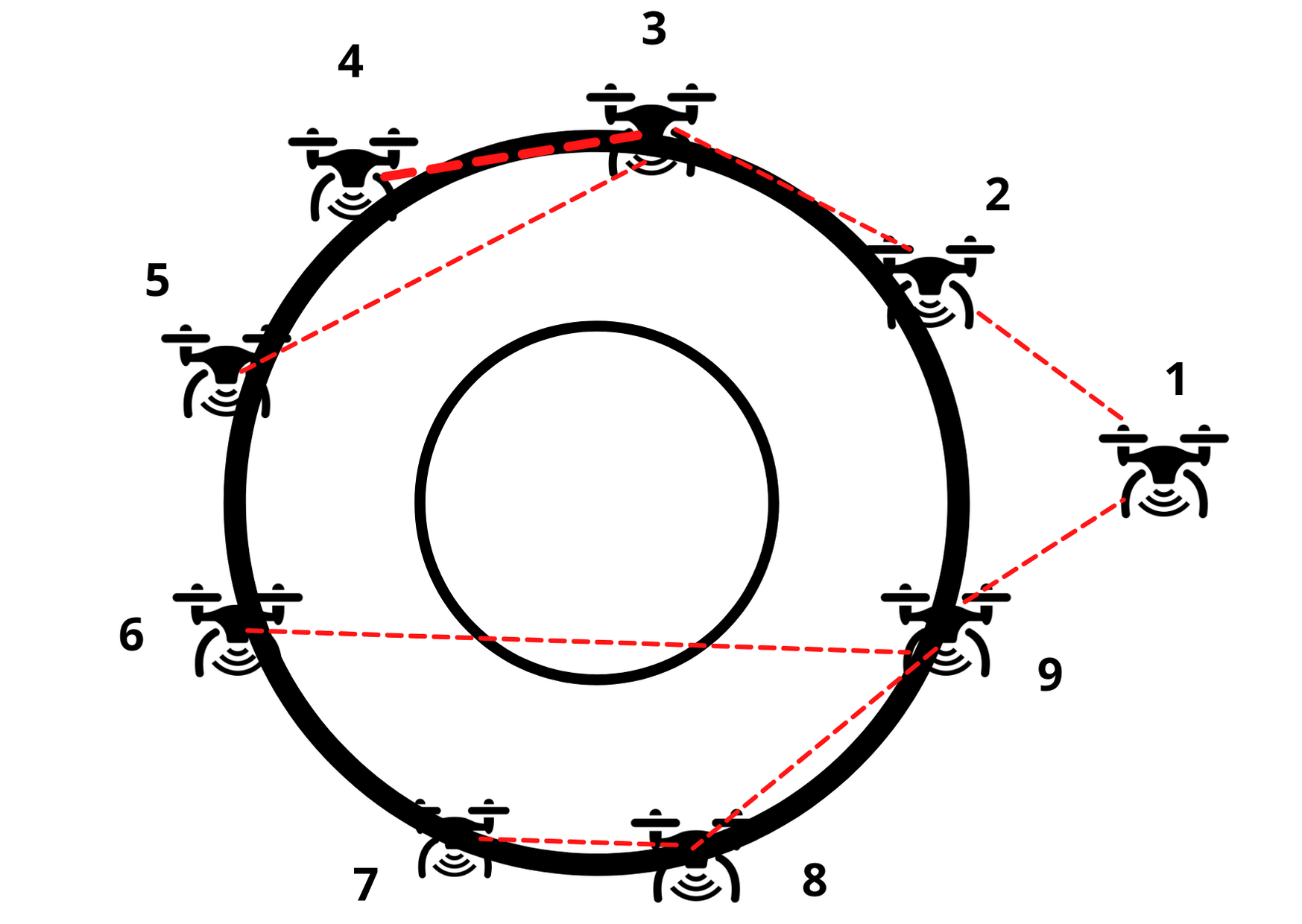}}
\caption{A possible edge formation between UAVs for multi hop communication}
\label{multi}
\end{figure}
\begin{figure}[htbp]
\centerline{\includegraphics[width=8cm]{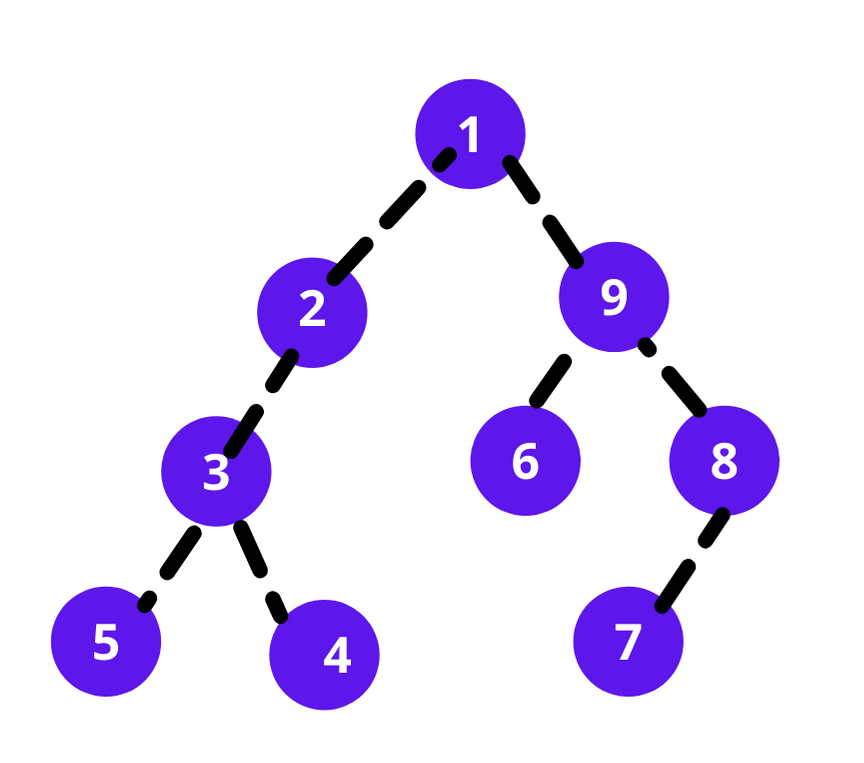}}
\caption{Spanning Tree for Fig. \ref{multi} }
\label{tree}
\end{figure}

From the figure \ref{multi}, the edges are formed considering distance and network load on the receiver UAV as weights. This way, as soon as the receiver gets the packet, it can forward to the next UAV. For the topology of as in Fig. \ref{multi}, a minimum spanning tree has been obtained as Fig. \ref{tree} using dijkstra's algorithm.
\section{Simulation Results}
We have developed a python based simulator to implement the above mentioned concepts to make the UAV assisted communication systems more efficient. The parameters considered for the simulation are radius of the circular topology $R' =200$, maximum range of UAV $R=40m$, number of Users $N=150$, altitude of flight $H=90 m$ and the maximum buffer at all the UAVs i.e., $B=50$. 
\subsection{Coverage Area}
\begin{figure}[htbp]
\centerline{\includegraphics[width=8cm]{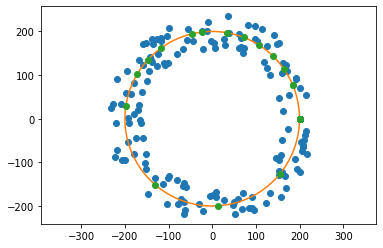}}
\caption{The optimal number of the system with traffic requirements into consideration. }
\label{optimaluav}
\end{figure}

The objective of obtaining maximum coverage area can be achieved by obtaining minimal number of UAVs at appropriate flying locations. The minimal number of uavs required according to the system parameters defined above is shown in the Fig. \ref{optimaluav}. SNR (Signal-to-Noise ratio) concept can be used to understand the coverage of the deployed UAVs. We have used Friis equation to obtain the basic path loss model because LoS link is extremely dominant in UAV assisted communications. So, fading, shadowing and delay spread problems of wireless communications can be avoided .

The modified Friis free space propagation in terms of area of transmitter and receiver can be shown as \eqref{friis}, where $\lambda$ is the wavelength ($10mm$ is considered for the simulation) and $d$ is the distance between the transmitter and receiver. Most of the literature considered the noise power to be in between $30$ dBm to $50$ dBm.

\begin{equation}
    P_r=\frac{A_t A_r}{ \lambda ^2 d^2}
    \label{friis}
\end{equation}

\begin{figure}[htbp]
\centerline{\includegraphics[width=8cm]{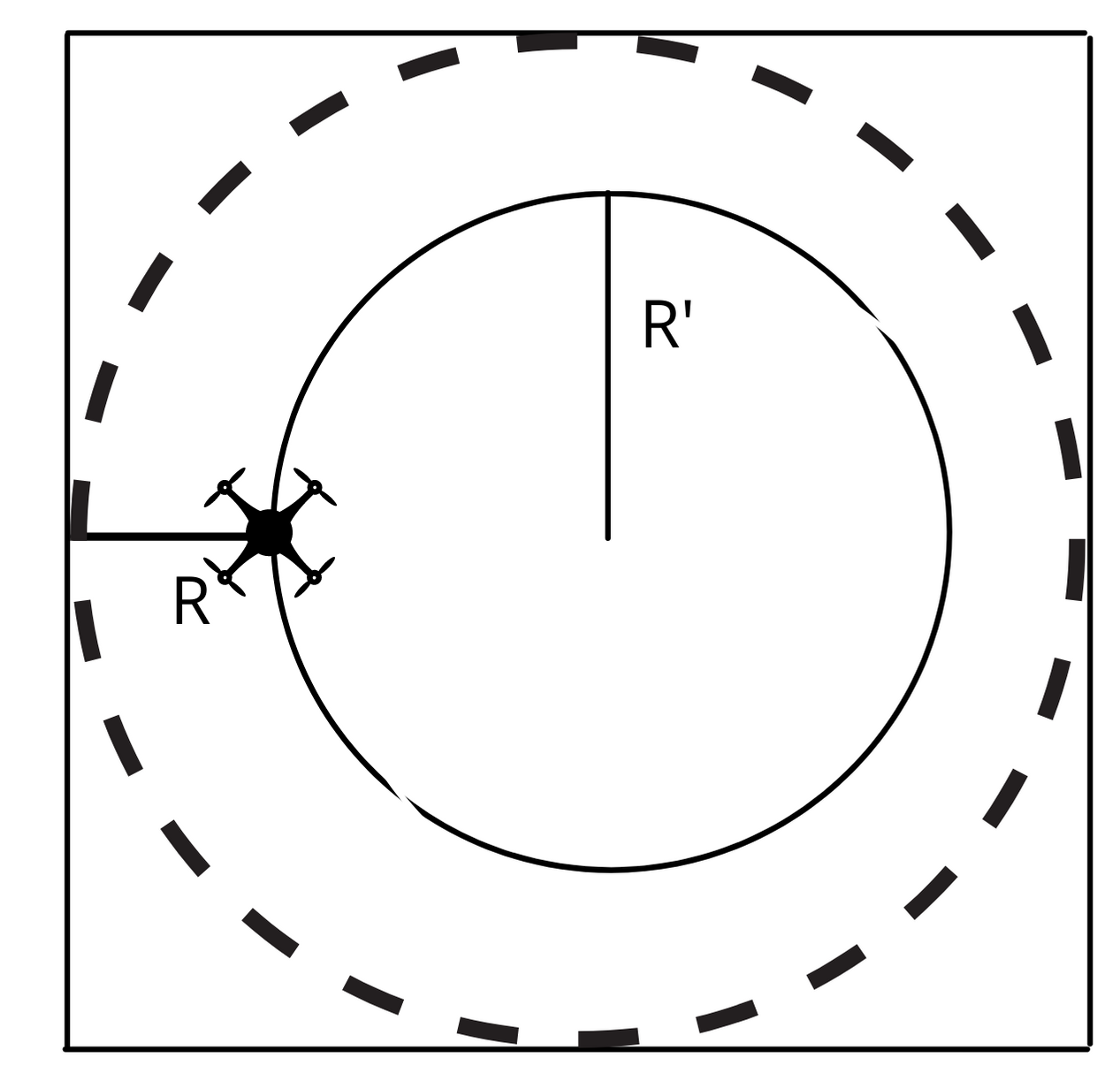}}
\caption{Computation of SNR inside the square is depicted.}
\label{square}
\end{figure}

\begin{figure}[htbp]
\centerline{\includegraphics[width=8cm]{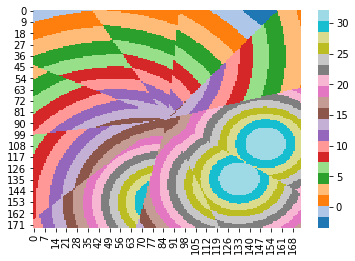}}
\caption{Heat map of the system with SNR at the initial deployment of UAVs .}
\label{FinalInit}
\end{figure}

\begin{figure}[htbp]
\centerline{\includegraphics[width=8cm]{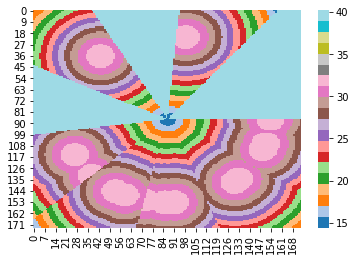}}
\caption{Heat map of the system with SNR at the optimal positioning of UAVs .}
\label{FinalOptim}
\end{figure}

As shown in the Fig. \ref{square}, we consider a square of side length $2(R+R')$ with a center at origin to compute SNR at each coordinate inside the square. Using the friis free space model and noise power as $30$ dBm we obtain a heatmap as shown in the Fig. \ref{FinalInit}. The heat map obtained is of the initial deployment of the UAVs by considering the traffic requirements in each segment and their maximum range ($R$) of service. In the obtain we can see concentric circles for few sectors, which indirectly depicts that only one UAV is deployed in that sector and multiple UAVs are deployed in rest of the circular area. In the initial deployment the distribution of SNR is from $-10$ dB to $30$ dB. After obtaining the minimal number of UAVs and placing them in the best possible locations, we obtain the heat map of the system as Fig. \ref{FinalOptim}. We can see more concentric circles obtained in Fig. \ref{FinalOptim} than in Fig. \ref{FinalInit} proving that the UAVs are situated in optimal locations as in Fig. \ref{optimaluav}. The minimum SNR required for effective communication is $14$ dB , at which the device can still function and deliver relatively optimal performance. The optimal scenario Fig. \ref{FinalOptim} has the SNR ranging from $15$ dB to $40$ dB. The SNR at the corners of the square i.e., beyond the range of UAV is $< 0$ .

\subsection{Traffic Congestion}
\begin{figure}[htbp]
\centerline{\includegraphics[width=8cm]{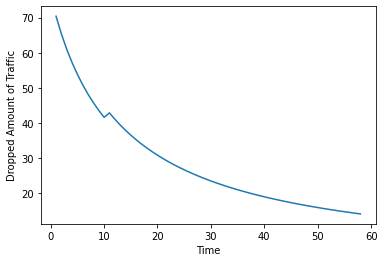}}
\caption{Plot for dropped amount of traffic vs time.}
\label{droppackets}
\end{figure}

The other main objective to improve the efficiency of the system is to reduce the number of packets dropped i.e., avoiding re-transmissions. To achieve this , mobility of the UAVs is utilized to obtain a better location of flight at each interval of time. We also implemented a traffic congestion algorithm i.e., feedback mechanism explained in above sections. Thus, the average amount of traffic dropped at all the UAVs with time is shown in fig. \ref{droppackets}, that satisfies our objective as mentioned in the section 4. At t = $11$ units,due to the surge in traffic, a new UAV is deployed in order to serve the traffic requirements. After the new deployment, the curve has become steeper and stable along with time. 

\section{Conclusion}
In this paper, we have investigated the key aspects of SDN assisted UAV communication to improve the efficiency of deployment. We considered a circular topology for deployment at stadiums, events and concerts. Then, we divide the circular area into various sectors depending on the UAV's maximum coverage range. Then, We proposed a heuristic to assign the UAVs to congested sectors and to evaluate each sector based on the amount of traffic and its priority i.e., real-time and non-realtime traffic. We implemented a traffic congestion control algorithm to avoid packet dropping as in \cite{7486987}. The power required for transmission is dependent on the distance between sender and receiver. Instead of direct communication between UAV and controller, we have explored multi-hop forwarding using Dijkstra's algorithm to find the best path from the sender UAV to the root UAV. Thus, the power consumption can be reduced to increase the flight time.

\section{Future Work}
\begin{figure}[htbp]
\centerline{\includegraphics[width=8cm]{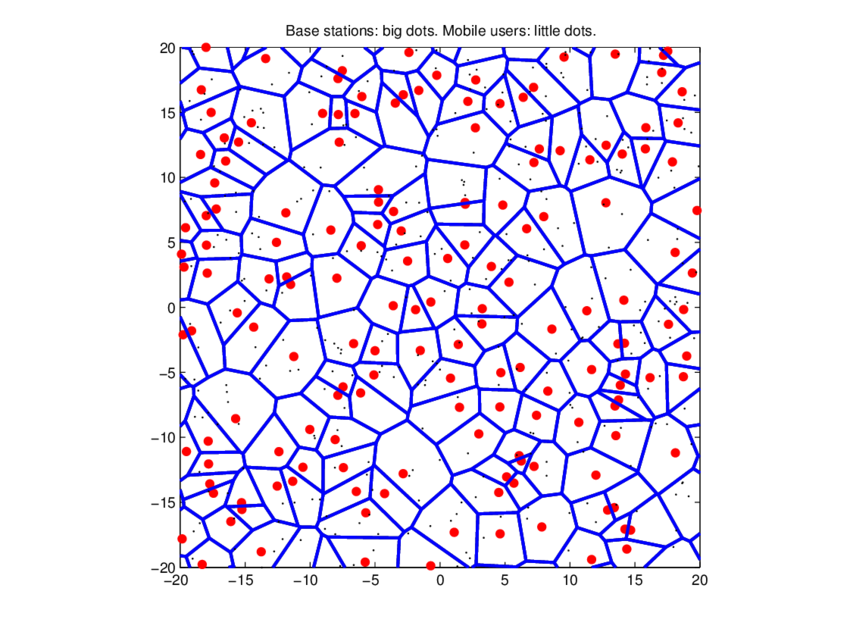}}
\caption{ Distribution of base stations and mobiles along with each mobile associated with the nearest BS using poisson distribution.
 }
\label{poisson}
\end{figure}

In this paper, we have considered a circular topology such as in stadiums,events and similar scenarios. A dynamic topology of UAVs must be obtained when mobility of the users is considered. In such cases finding the best path from UAV to root node can become more difficult because of the network load of the receiver UAV, distance between them, user-UAV traffic generation rate and UAV-user traffic generation rate. To simulate such real life scenarios considering all the above said aspects and various possible attenuations  can be extremely difficult (Fig .\ref{poisson}). But there is a good scope in the growth of UAV communications for their versatility and applications as discussed in this paper.

\bibliographystyle{IEEEtran}
\bibliography{ms}

\begin{thebibliography}{1}
\providecommand{\url}[1]{#1}
\csname url@samestyle\endcsname
\providecommand{\newblock}{\relax}
\providecommand{\bibinfo}[2]{#2}
\providecommand{\BIBentrySTDinterwordspacing}{\spaceskip=0pt\relax}
\providecommand{\BIBentryALTinterwordstretchfactor}{4}
\providecommand{\BIBentryALTinterwordspacing}{\spaceskip=\fontdimen2\font plus
\BIBentryALTinterwordstretchfactor\fontdimen3\font minus
  \fontdimen4\font\relax}
\providecommand{\BIBforeignlanguage}[2]{{%
\expandafter\ifx\csname l@#1\endcsname\relax
\typeout{** WARNING: IEEEtran.bst: No hyphenation pattern has been}%
\typeout{** loaded for the language `#1'. Using the pattern for}%
\typeout{** the default language instead.}%
\else
\language=\csname l@#1\endcsname
\fi
#2}}
\providecommand{\BIBdecl}{\relax}
\BIBdecl

\bibitem{ZOU2020}
\BIBentryALTinterwordspacing
C.~Zou, X.~Li, X.~Liu, and M.~Zhang, ``3d placement of unmanned aerial vehicles
  and partially overlapped channel assignment for throughput maximization,''
  \emph{Digital Communications and Networks}, 2020. [Online]. Available:
  \url{https://www.sciencedirect.com/science/article/pii/S2352864820302479}
\BIBentrySTDinterwordspacing

\bibitem{8038869}
M.~{Mozaffari}, W.~{Saad}, M.~{Bennis}, and M.~{Debbah}, ``Mobile unmanned
  aerial vehicles (uavs) for energy-efficient internet of things
  communications,'' \emph{IEEE Transactions on Wireless Communications},
  vol.~16, no.~11, pp. 7574--7589, 2017.

\bibitem{8450437}
H.~{Shakhatreh} and A.~{Khreishah}, ``Optimal placement of a uav to maximize
  the lifetime of wireless devices,'' in \emph{2018 14th International Wireless
  Communications Mobile Computing Conference (IWCMC)}, 2018, pp. 1225--1230.

\bibitem{9120678}
W.~{Ding}, Z.~{Yang}, M.~{Chen}, J.~{Hou}, and M.~{Shikh-Bahaei}, ``Resource
  allocation for uav assisted wireless networks with qos constraints,'' in
  \emph{2020 IEEE Wireless Communications and Networking Conference (WCNC)},
  2020, pp. 1--7.

\bibitem{9014076}
M.~D. {Nguyen}, T.~M. {Ho}, L.~B. {Le}, and A.~{Girard}, ``Uav placement and
  bandwidth allocation for uav based wireless networks,'' in \emph{2019 IEEE
  Global Communications Conference (GLOBECOM)}, 2019, pp. 1--6.

\bibitem{8190924}
S.~{ur Rahman}, G.~{Kim}, Y.~{Cho}, and A.~{Khan}, ``Deployment of an sdn-based
  uav network: Controller placement and tradeoff between control overhead and
  delay,'' in \emph{2017 International Conference on Information and
  Communication Technology Convergence (ICTC)}, 2017, pp. 1290--1292.

\bibitem{7569080}
V.~{Sharma}, R.~{Sabatini}, and S.~{Ramasamy}, ``Uavs assisted delay
  optimization in heterogeneous wireless networks,'' \emph{IEEE Communications
  Letters}, vol.~20, no.~12, pp. 2526--2529, 2016.

\bibitem{7486987}
M.~{Mozaffari}, W.~{Saad}, M.~{Bennis}, and M.~{Debbah}, ``Efficient deployment
  of multiple unmanned aerial vehicles for optimal wireless coverage,''
  \emph{IEEE Communications Letters}, vol.~20, no.~8, pp. 1647--1650, 2016.

\end{thebibliography}

\end{document}